\documentclass[a4paper,11pt]{article}
\usepackage[numbers,sort&compress]{natbib}
\usepackage{jheppub}

\usepackage{longtable}
\usepackage{physics}
\usepackage{ytableau}
\ytableausetup{smalltableaux}
\usepackage{amsmath}
\usepackage{amsfonts}
\usepackage{mathrsfs}
\usepackage{float}
\usepackage{stackengine}
\usepackage{amssymb}
\usepackage{amsthm}
\usepackage{hyperref}
\usepackage{mathtools}
\usepackage{bbm}
\usepackage[dvipsnames]{xcolor}
\usepackage{dsfont}
\usepackage{graphicx}
\graphicspath{ {images/} }

\usepackage[T1]{fontenc}

\title{Regulating Free-scalar and Yang--Lee 5d CFTs on the Fuzzy Four Sphere}
\author[a]{Jiangyuan Qian}
\affiliation[a]{Department of Particle Physics and Astrophysics, Weizmann Institute of Science, Rehovot 7610001, Israel}
\emailAdd{jiangyuan.2718@gmail.com}

\abstract{Non-commutative sphere regularization of CFT has become a powerful tool for computing conformal data of 3d CFTs and was recently extended to 4d CFTs. In this work we further construct 5d free-scalar CFT and Yang--Lee CFT on a fuzzy 4-sphere. We observe a continuous phase transition and conformal towers in these examples. We obtain the first three scalar primaries of the 5d free-scalar CFT, furthermore, we compute the 5d Yang--Lee $\Delta_\phi,\Delta_{\phi^3}$ with $N_e=10$ fermions using exact diagonalization (ED) and $N_e=20$ fermions using density matrix renormalization group (DMRG), obtaining relatively good agreement with both $\epsilon$-expansion and the Gliozzi truncated bootstrap result.}
\keywords{Field Theories in Higher Dimensions, Scale and Conformal Symmetries}

\begin{document}

\maketitle
\flushbottom

\section{Introduction}
Conformal Field Theories(CFTs) are a central topic in the study of high energy physics. The symmetry structure restricts the form of observables in the theory making it solvable non-perturbatively after determining a small set of conformal data. However, solving the conformal data $(\Delta_i,C_{ij}^k)$ of CFT is a hard problem, since many conformal field theories with small numbers of degrees of freedom are strongly coupled. In some cases, there are perturbative techniques such as the $\epsilon$-expansion\cite{eps_exp}, and the large N expansion\cite{1_by_N_exp}, and non perturbative techniques such as conformal bootstrap\cite{CB,El-Showk_12}.

The fuzzy sphere regularization is a new method\cite{Wei_23,Han_23} which computes the conformal data for a 3d CFT with these steps:
\begin{itemize}
    \item Find a phase transition that is described by the target CFT.
    \item Put it on $\mathbb S^2\times \mathbb R$, and do the Lowest Laudau Level projection, which contains a finite number of orbitals.
    \item Tune to the critical point, by matching the finite size scaling or conformal tower to CFT.
    \item Compute the spectrum $E_n-E_0\simeq \frac{v \Delta_n}{R}$, and $\langle i|\mathcal O_j|k\rangle\propto C_{ikj}$.
\end{itemize}

The Lowest Landau Level(LLL) projection requires fermionic or bosonic particles that are charged under a topological winding gauge field. For fuzzy 2-sphere, one considers Schr\"odinger particles with mass M and charged under a $\mathrm{U}(1)$ background gauge field with magnetic flux $\int_{\mathbb S^2} dA=4\pi s$. The single particle energy levels are: $E_n =\frac{n(n+1)+(2n+1)s}{2Mr^2}$, with $2s+2n+1$ fold degeneracy forming a $j=s+n$ representation of the $\mathrm{SO}(3)$ rotation symmetry. By taking particle mass $M\rightarrow0$ limit, the physical Hilbert space (which is a many-body Hilbert space) is constructed solely from $n=0$ orbitals (LLL orbitals), and this provides a method to regularize the UV divergence. This method was shown to be valid and powerful in computing the conformal data for many strongly coupled CFTs in 3d, such as $O(N)$ Wilson Fisher CFTs and Chern-Simons matter CFTs \cite{taylor_25,Yang_25,Guo_25,dey_26,dey_26_2,Yang_25_DQCP}.

It is natural to extend this method to a fuzzy 4-sphere, by considering mass $M$ Schr\"odinger particles charged with spin $I/2$ under an $\mathrm{SU}(2)$ gauge field with an $\mathrm{SO}(5)$-invariant instanton background on $\mathbb S^4\times \mathbb R$\cite{Yang_77qv,Yang_78td,Hasebe_20,Zhang_01}, in the $M\rightarrow0$ limit. There are $d(I,0)={I+3\choose3}$ degenerate orbitals which are analogous to the fuzzy 2-sphere LLL orbitals. A recent numerical work\cite{Meng_26} shows that if one further dimensionally reduces it to $\mathbb S^3$, it properly reproduces the spectrum of 4d free-scalar CFT and Yang--Lee CFT, which sheds light on extending this method to higher dimensional CFTs.

In this work we apply this method to 5d CFTs, by constructing the critical Ising model on a fuzzy 4-sphere. As the $\mathbb Z_2$ universality class has upper critical dimension $D_c=4$, we expect a free-scalar CFT. Numerically for $I=2$ we obtain a continuous phase transition and conformal spectrum of the 5d free-scalar CFT on $\mathbb S^4 \times\mathbb R$, at least for symmetric tensor representations with small spin.

We further construct the 5d Yang--Lee model by adding a $\mathcal R$ symmetry ($\phi\leftrightarrow-\phi$) breaking term to the Ising model while keeping the $\mathcal R \mathcal T$ symmetry unbroken, similar to the low dimension Yang--Lee model. When increasing the magnitude of the $\mathcal R$ symmetry breaking term $h_z$, real energy levels merge into complex conjugate pairs, which is a continuous transition towards an $\mathcal R \mathcal T$-broken phase. The critical point is described by a CFT with potential $i\phi^3$, which is below its upper critical dimension $D_c=6$. We compute the Yang--Lee scaling dimension for $I=2,3$, finding $\Delta_\phi =1.44$ and $\Delta_{\phi^3}=5.75$, with a good agreement to the $\epsilon$-expansion and to the conformal bootstrap results.

In section 2, we will write down a generic $\mathrm{SO}(5)$ symmetric four-fermion interaction and impose constraints from locality. In section 3, we will construct the Ising phase transition and find the critical point by matching the integer level spacing of primary and descendants. We indeed find a critical point exhibiting free-scalar CFT spectrum. In section 4, we construct a model with $\mathcal R \mathcal T$ symmetry breaking transition and find a critical point and critical Yang--Lee spectrum. We further compute the scaling dimension of $\phi,\phi^3$ by ED in $I=2,N_e=10$ and by DMRG in $I=3,N_e=20$.

\section{Fermion interactions}
One of the ways to construct a fuzzy 4-sphere is to consider an $\mathrm{SU}(2)$ instanton on $\mathbb S^4\times\mathbb R$, which satisfies $\frac{1}{24}\int_{\mathbb S^4} \text{tr}(F\wedge F) =\frac{8\pi^2}{3}c_2$. The $\mathrm{SO}(5)$ symmetric solution (with positive $c_2$) is unique and with instanton number $c_2=1$, and one can construct a fuzzy sphere with different radius by taking a Schr\"odinger particle in the spin $I/2$ representation of an $\mathrm{SU}(2)$ gauge group coupled to this gauge field. The energy levels are quantized and the single particle wave functions transform in the $(I+n,n)_5$ representation of $\mathfrak{so}(5)$ (as $\mathfrak{sp}(4)\simeq \mathfrak{so}(5)$ we will use the two notations interchangeably, all Dynkin labels in this paper are written in $\mathfrak{sp}(4)$ notation, see details in appendix \ref{app: sp4_rep}). If further take the $M\rightarrow0$ limit (physically take $1/Mr^2$ larger than any interaction energy scale), the single particle state is in the $(I,0)_5$ representation\footnote{A symmetric product of $I$ spinors of $\mathfrak{so}(5)$.} with degeneracy $d(I,0)={I+3\choose3}$.

In the Dirac monopole construction of a fuzzy 2-sphere, the four-fermion interaction reads:
\begin{equation}
    V_{m_1,m_2,m_3,m_4}=\sum_{l=0}^{2s}(4s-2l+1) V_{l} \begin{pmatrix}
        s&s& 2s-l \\
        m_1& m_2 &-m_1-m_2
    \end{pmatrix}\begin{pmatrix}
        s&s& 2s-l \\
        m_4& m_3 &-m_3-m_4
    \end{pmatrix},\label{eq: psedu-potential}
\end{equation}
and for a short ranged interaction, $V_l$\footnote{Describing the two spin $s$ representations joining into a representation of spin $2s-l$.} are only non-zero for a finite number of $l=0,1,...,k$ for arbitrarily large $s$. We can understand this result in an intuitive picture. Consider fermions localized at the north pole (we also assume there are some flavor indices) of the sphere, they both occupy the $m=s$ orbital, so this fermion pair sits in the representation with spin $2s$ and can feel a local interaction. Thus one needs to open the $l=0$ channel for a local interaction. We would like to generalize it to fuzzy $\mathbb S^4$, where the LLL projection and local interactions preserve $\mathrm{SO}(5)$ symmetry. 
After doing the mode expansion $\Psi(\Omega_4)= \frac{1}{R^2} \sum_{i}\Psi_i Y_{i}^{(I,0)}(\Omega_4)$\footnote{Previously, $r$ we denote as the physical radius, while $R$ is dimensionless and can be intuitively as $R(I)=r/l_{UV}$.} where the index $i$ goes over the components of the $(I,0)_5$ representation, the generic form of interaction is given by:
\begin{equation}
  V_{\text{int}}=  \sum_{i_1,i_2,i_3,i_4}V_{i_1,i_2,i_3,i_4} \Psi^\dagger_{i_1} \Psi^\dagger_{i_2}\Psi_{i_3}\Psi_{i_4},
\end{equation}
where the $\mathrm{SO}(5)$ invariance ensures:
\begin{equation}
    \sum_{j_1,j_2,j_3,j_4}V_{j_1,j_2,j_3,j_4}D_{j_1i_1}^*(g) D_{j_2i_2}^*(g) D_{j_3i_3}(g) D_{j_4i_4}(g)= V_{i_1,i_2,i_3,i_4}.
\end{equation}
If we write $(i_1,i_2)=\mathrm A$, $(j_1,j_2)=\mathrm A'$ and $(i_3,i_4)=\mathrm B$, $(j_3,j_4)=\mathrm B'$ as integrated labels, we could write the above equation in a matrix form: $V_{\mathrm A'\mathrm B'}D_{\mathrm A'\mathrm A}^*D_{\mathrm B'\mathrm B}=V_{\mathrm A\mathrm B}$. As $D$ is a unitary matrix, $\mathbf V \mathbf D =\mathbf D \mathbf V$, from Shur's lemma, $\mathrm{SO}(5)$ invariance constraints $V_{\mathrm A\mathrm A'}$ to be a block diagonal matrix in each irreducible representation. As the integrated label $\mathrm A$ labels the states in $(I,0)_5\otimes(I,0)_5$ representation, the form of $V_{\mathrm A\mathrm A'}$ is specified by assigning a pseudo-potential $V_{r,s}$ describing the two $(I,0)_5$ fermions joining into a representation $(2I-2s-2r,r)_5$, and thus the fermion interaction matrix is given by:
\begin{equation}
    V_{i_1,i_2,i_3,i_4} =\sum_{r\geq0,s\geq0}^{r+s\leq I}\sum_{i=1}^{d(2I-2r-2s,r)} V_{r,s}C_{i_3i_4}^{(2I-2r-2s,r,i)}C_{i_1i_2}^{(2I-2r-2s,r,i)*}.
\end{equation}
This is essentially the same as the pseudo-potential parameterization (\ref{eq: psedu-potential}) in the fuzzy 2-sphere, and we need to subsequently compute the $\mathfrak so(5)\simeq \mathfrak{sp}(4)$ Clebsch-Gordan(CG) coefficients. Instead of using a lengthy expression in \cite{Holman_73,Hongoh_74}, we use an algorithm in appendix \ref{app: sp4_rep} to obtain $C_{\mathbf N_a \mathbf N_b}^{(p,q,i)}$ numerically in the Schwinger boson basis. In the following text, we use an occupation vector $\mathbf N_a=(n_a^1,n_a^2,n_a^3,n_a^4)$ satisfying $\sum_{m=1}^4n_a^m=I$ to replace the basis label $i_a$ for the LLL orbitals. Similarly we will use $V_{\mathbf{N_1},\mathbf{N_2},\mathbf{N_3},\mathbf{N_4}}$ as the four-fermion interaction matrix, and we list some of its properties:
\begin{equation}
    \begin{split}
V_{\mathbf{N_1},\mathbf{N_2},\mathbf{N_3},\mathbf{N_4}}&= V_{\mathbf{N_3},\mathbf{N_4},\mathbf{N_1},\mathbf{N_2}},\\
        V_{\mathbf{N_1},\mathbf{N_2},\mathbf{N_3},\mathbf{N_4}}&=V_{\mathbf{N_2},\mathbf{N_1},\mathbf{N_4},\mathbf{N_3}},\\
V_{\mathbf{N_1},\mathbf{N_2},\mathbf{N_3},\mathbf{N_4}}&=V_{\mathbf{N_1},\mathbf{N_2},\mathbf{N_3},\mathbf{N_4}}^*.
    \end{split}
\end{equation}
The algorithm in appendix \ref{app: sp4_rep} implies that all the $C_{\mathbf N_1, \mathbf N_2}^{(p,q,i)}$ can be chosen real, which leads to the first line. From the symmetry of the product state, the CG coefficients satisfy $C_{\mathbf N_1, \mathbf N_2}^{(2I-2r-2s,r,i)}=C_{\mathbf N_2, \mathbf N_1}^{(2I-2r-2s,r,i)}(-1)^{r}$, leading to the second line. In this paper we take $V_{r,s}$ to be real, which gives the third line.

\begin{figure}
    \centering
    \includegraphics[width=0.8\linewidth]{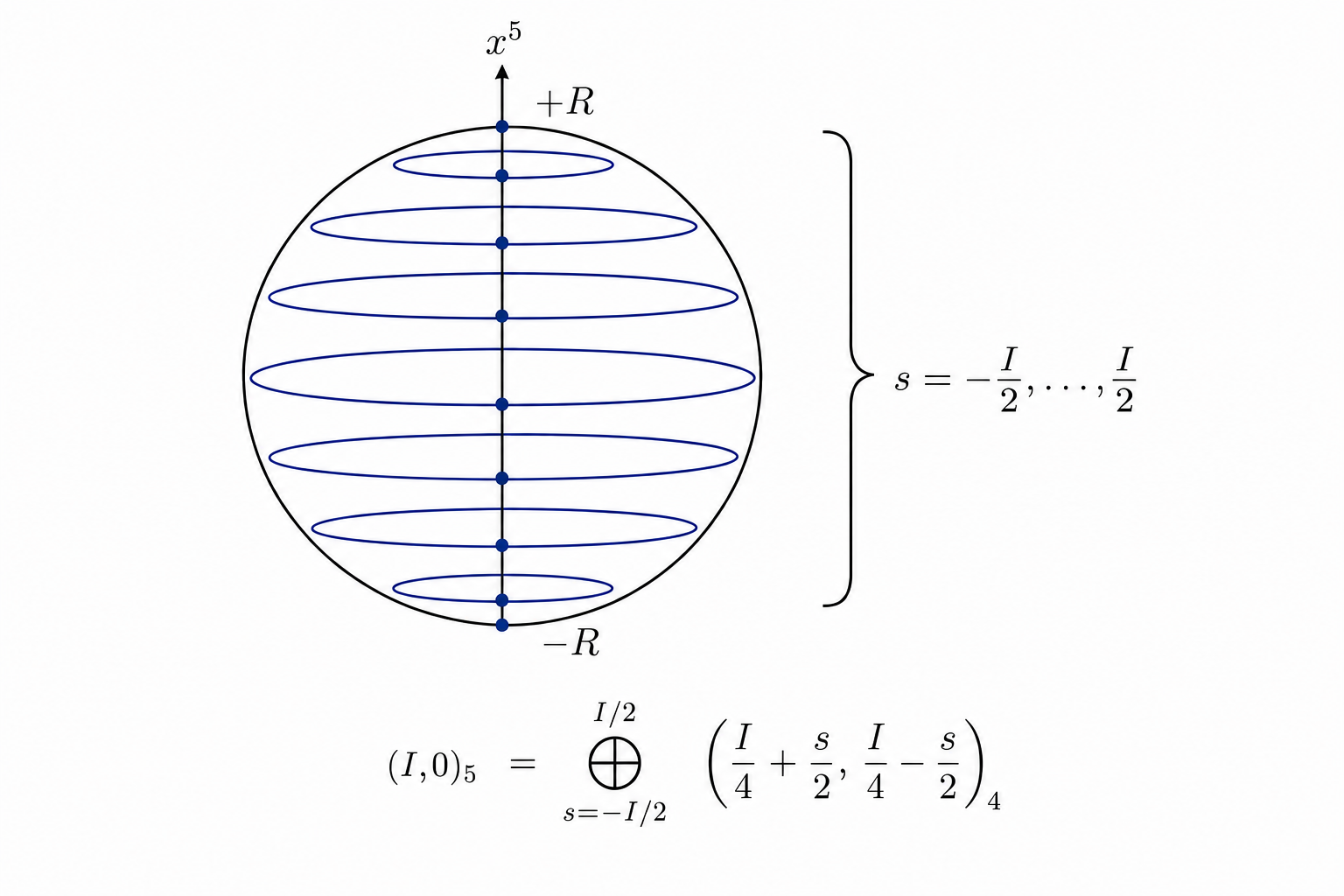}
    \caption{The fuzzy $\mathbb S^4$ orbital decompose into $\mathbb S^3$ orbitals.}
    \label{fig:fuzzy S4 decompose}
\end{figure}
Which channel $V_{r,s}$ is opened for a UV local Hamiltonian is a question not yet answered. In this paper, we set the pseudo-potentials $V_{0,0},V_{1,0},V_{0,1}$ to be non-zero, as the first few local terms in the gradient expansion, if take a pair of fermions at the north pole in the $x^5$ direction, they transform under the $(I/2,0)_4\otimes(I/2,0)_4$ representation of the $\mathfrak{so}(4)$ symmetry rotating the $\mathbb S^3$ around this point. As the maximally polarized pair state on $\mathbb S^3$ has interactions (from the locality of the interaction), we need to open the $(I,0)_4$ representation which is only contained in $(2I,0)_5$ among $(2I - 2r-2s, s)_5$.

The next order of the gradient expansion contains terms either adjacent in the $x^5$ direction (see figure \ref{fig:fuzzy S4 decompose} for the branching of $\mathbb S^4$ orbitals to $\mathbb S^3$ in $x^5$ direction) $(I-1/2,1/2)_4\subset (I/2,0)_4\otimes(I/2-1/2,1/2)_4$, or adjacent in $\mathbb S^3$ direction, $(I-1,0)_4\subset(I/2,0)_4\otimes(I/2,0)_4$. $(I-1/2,1/2)_4$ is contained in $(2I,0)_5, (2I-2,1)_5$, while $(I-1,0)_4$ is contained in $(2I-2,1)_5$ and $(2I-2, 0)_5$, so we propose these are the channels that need to be opened.

\section{Free-scalar CFT}
\begin{figure}[t]
    \centering
    \begin{minipage}[t]{0.45\linewidth}
    \stackinset{l}{2mm}{t}{2mm}{\textbf{a.}}{
        \includegraphics[width=\linewidth]{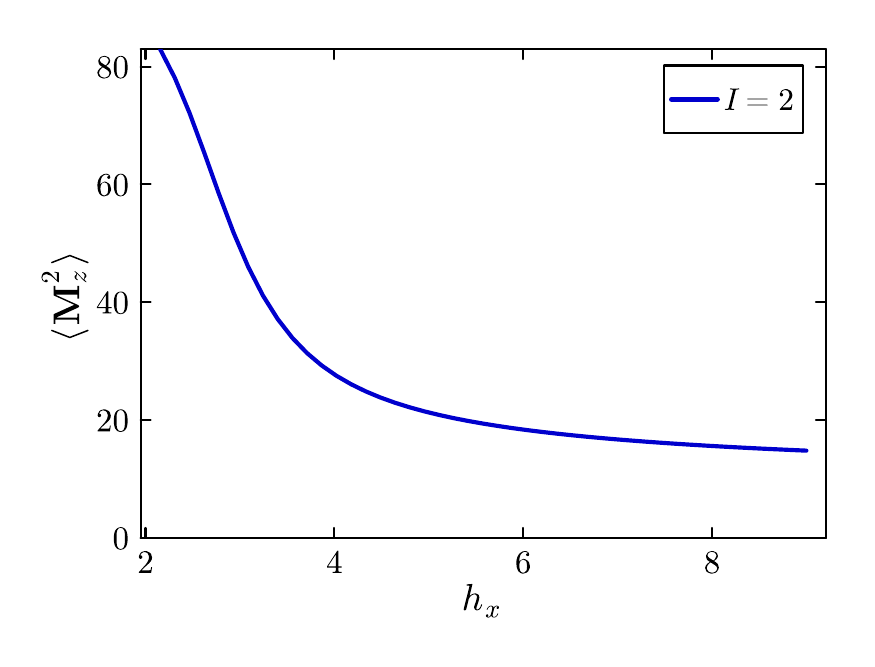}
        }
    \end{minipage} 
    \begin{minipage}[t]{0.45\linewidth}
    \stackinset{l}{2mm}{t}{2mm}{\textbf{b.}}{
        \includegraphics[width=\linewidth]{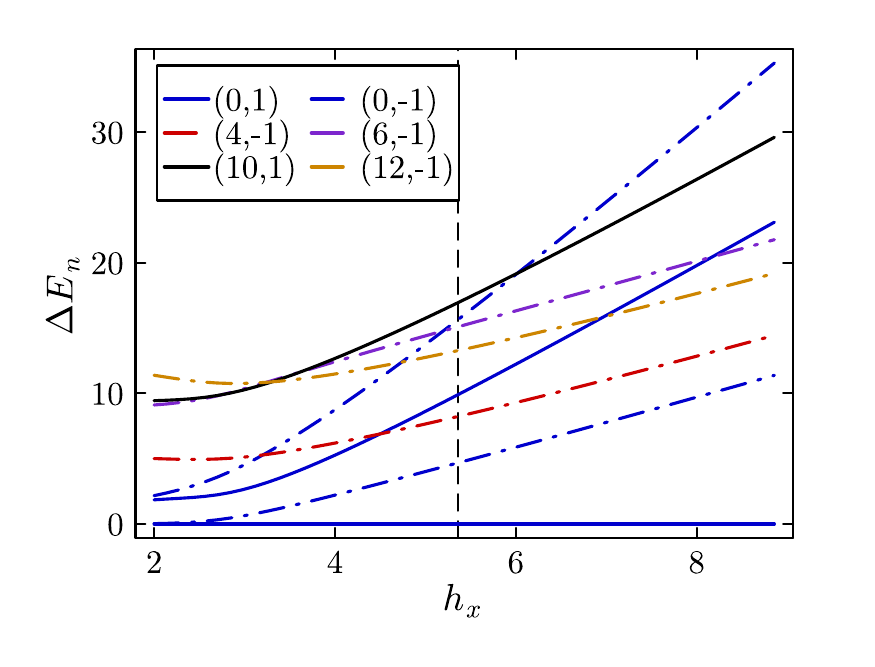}
        }
    \end{minipage}
    \caption{(a) $\langle\mathbf{M_z}^2\rangle$ as a function of $h_x$, for $V_{1,0}=1.600, V_{0,1}=0.011$. (b) Energy gaps in different $(C_2, \mathcal R, +1)$ sectors as a function of $h_x$, with same parameter choice.}
    \label{fig:Mz2 vs h}
\end{figure}

\begin{figure}[t]
    \centering
    \begin{minipage}[t]{0.45\linewidth}
    \stackinset{l}{2mm}{t}{2mm}{\textbf{a.}}{
        \includegraphics[width=\linewidth]{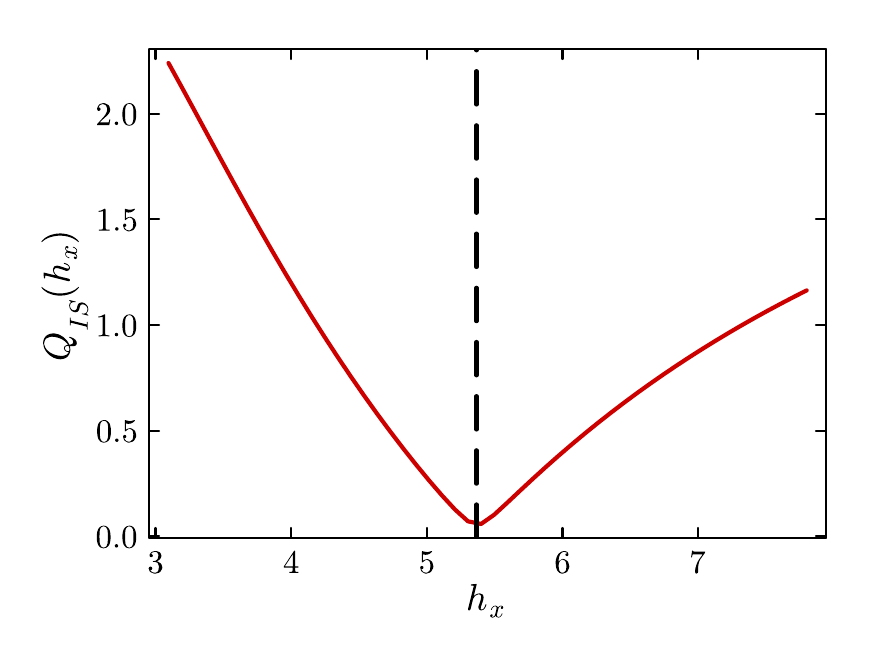}
        }
    \end{minipage}
    \begin{minipage}[t]{0.45\linewidth}
    \stackinset{l}{2mm}{t}{2mm}{\textbf{b.}}{
        \includegraphics[width=\linewidth]{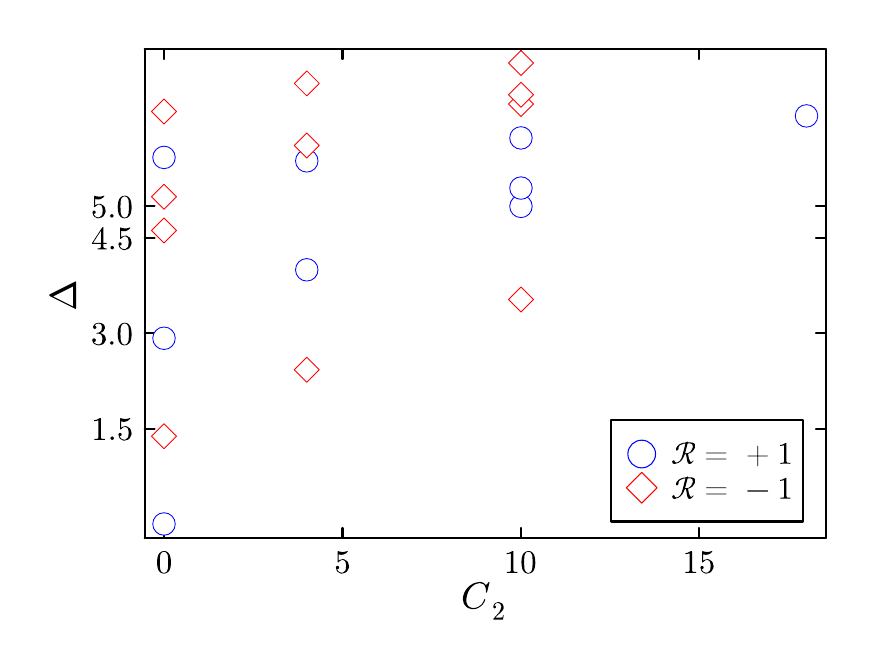}
        }
    \end{minipage}
        \begin{minipage}[t]{0.8\linewidth}
    \stackinset{l}{2mm}{t}{2mm}{\textbf{c.}}{
        \includegraphics[width=\linewidth]{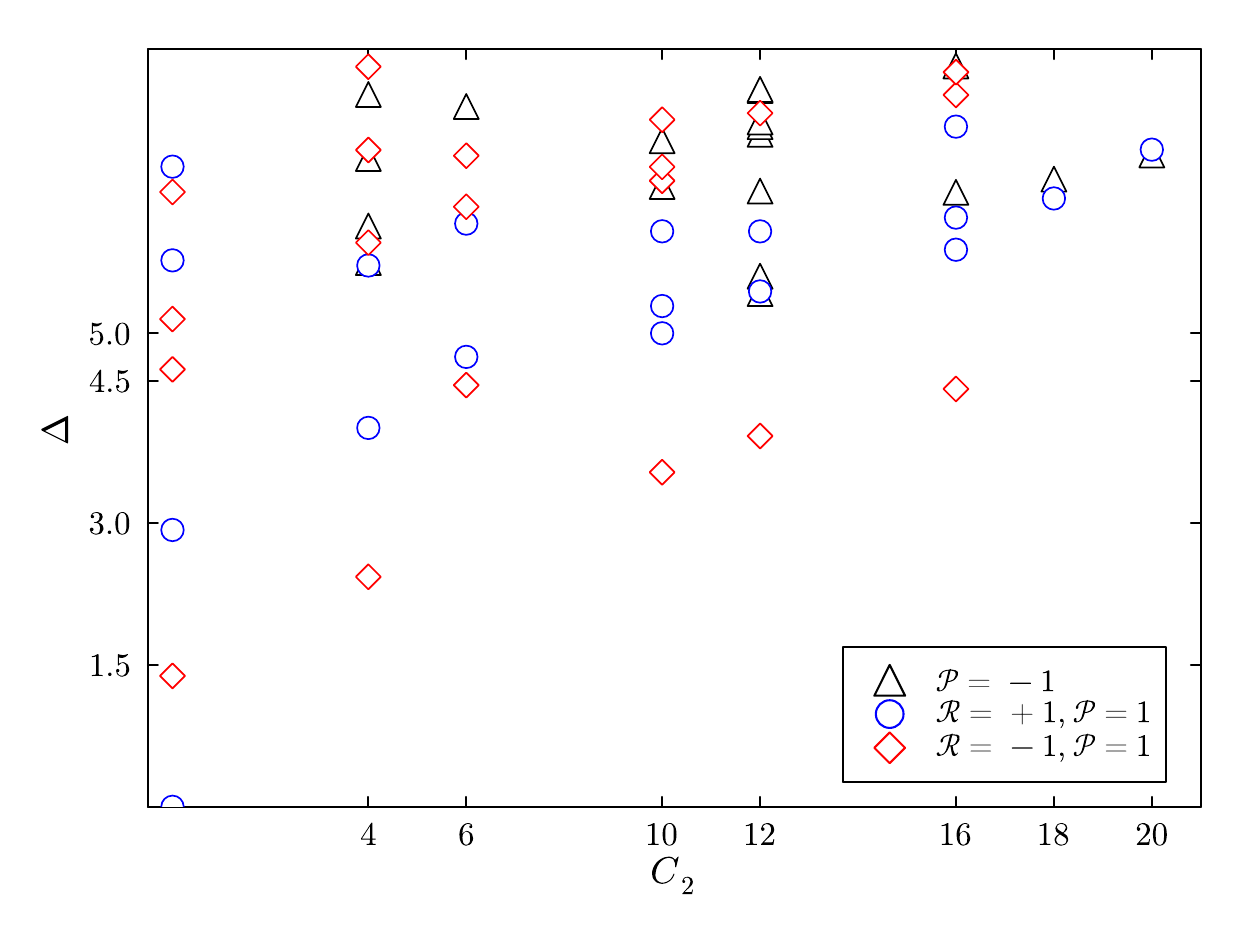}
        }
    \end{minipage} 
    \caption{(a) The Ising cost $Q_{IS}$ as a function of $h_x$, for $V_{1,0}=1.600, V_{0,1}=0.011$.  (b) The spectrum of $\mathcal P=+1$ symmetric tensor representations of $\mathfrak{so}(5)$, having removed the weight degeneracy in an irreducible representation. (c) The full spectrum. (b--c) are computed at $I=2$, with the parameters in (\ref{eq:Ising_para_ED}).}
    \label{fig:spec_IS}
\end{figure}

We take two fermions in the $(I,0)_5$ representation as above, and consider an Ising model with Hamiltonian:
\begin{equation}
\begin{split}
    H &= \sum_{\mathbf{N_1},\mathbf{N_2},\mathbf{N_3},\mathbf{N_4}} V_{\mathbf{N_1},\mathbf{N_2},\mathbf{N_3},\mathbf{N_4}}(::\Psi^\dagger_{\mathbf{N_1}}\Psi_{\mathbf{N_4}}\Psi^\dagger_{\mathbf{N_2}}\Psi_{\mathbf{N_3}}::+::\Psi^\dagger_{\mathbf{N_1}}\sigma^z\Psi_{\mathbf{N_4}}\Psi^\dagger_{\mathbf{N_2}}\sigma^z\Psi_{\mathbf{N_3}}::)\\
    &- h_x \sum_{\mathbf{N}}\Psi^\dagger_{\mathbf N} \sigma^x\Psi_{\mathbf N},
\end{split}
\end{equation}
analogously to the fuzzy $\mathbb S^2$ case reviewed above, where the Schwinger boson vector $\mathbf{N}_a=(n_a^1, n_a^2, n_a^3, n_a^4)$ labels the $(I,0)_5$ representation, with constraint $\sum_{m=1}^4n_a^m =I$, $\Psi_{\mathbf N} =(\psi_{\mathbf N,\uparrow},\psi_{\mathbf N,\downarrow})^T$, we fill ${I+3\choose3}$ out of $2{I+3\choose3}$ orbitals, the number of particles is $N_e=\frac{(I+3)(I+2)(I+1)}{6}$. The discrete symmetry of the above Hamiltonian is written as:
\begin{equation}
\begin{split}
   \mathcal R&: \Psi_{\mathbf N}\rightarrow\sigma^x \Psi_{\mathbf N}\\
   \mathcal P&: \Psi_{\mathbf N}\rightarrow i\sigma^y \Psi_{\mathbf{N}}^\dagger,
\end{split}
\end{equation}
where to see the $\mathcal P$ symmetry one needs to use $V_{\mathbf N_1, \mathbf N_2,\mathbf N_3,\mathbf N_4}=V_{\mathbf N_4, \mathbf N_3,\mathbf N_2,\mathbf N_1}$. The ferromagnetic order parameter is defined by:
\begin{equation}
    \mathbf M _z = \sum_{\mathbf N,\sum n^m=I}\mathbf \Psi ^\dagger_{\mathbf N} \sigma^z  \mathbf\Psi _{\mathbf N}.
\end{equation}
When tuning $h_x$ from small to large value, $\langle\mathbf M_z^2\rangle$ exhibit a smooth transition, see figure \ref{fig:Mz2 vs h}(a). At small $h_x$, the energy splitting of the $\mathcal R$ even and $\mathcal R$ odd ground state is small compared to higher excitation states, and it grows large at large $h_x$, see figure \ref{fig:Mz2 vs h}(b). These establish a continuous phase transition from ferromagnetic phase to paramagnetic phase.
As expected in the state operator correspondence: $E_n-E_0\simeq \frac{v\Delta_n}{R}$, we normalize the spectrum by subtracting the ground state energy and setting the first state\footnote{In 5d free-scalar CFT there is degeneracy in this sector with $T_{\mu\nu}$ and $\partial_\mu \partial_\nu \phi^2$, one may identify the first or the second state as the $T_{\mu\nu}$ in numerics.} in the ($C_2, \mathcal R, \mathcal P$)=($10,+,+$) sector as $\Delta_T=5$, $C_2$ is the Casimir of $\mathfrak{sp}(4)$, in table \ref{tab:sp4 reps} we list $C_2$ values of $\mathfrak{sp}(4)$ representations of form $(2n,q)_5$\footnote{LLL projection at $\nu=1$ the Hilbert space comes from decomposing $\frac{(I+1)(I+2)(I+3)}{6}$ copy of $(I,0)$ state, thus the central representation is $(-1)^{\frac{I(I+1)(I+2)(I+3)}{6}}=1$, the many-body states restricted to $(2n,q)$ states with center quantum number $+1$.}, to be specific, the symmetric rank-2 tensor sits in $C_2=10$. We obtain the parameters $V_{1,0}, V_{0,1},h_x$ by minimizing the cost function:

\begin{table}[t]
    \centering
    \setlength{\tabcolsep}{12pt}
    \renewcommand{\arraystretch}{1.3}
    \begin{tabular}{ccccc}
        \hline
       $(p,q)_5$  & $C_2$ & (0,0) degeneracy & \text{dimension} & \text{Tensor rep. of $\mathfrak{so}(5)$} \\
       \hline
        (0,1) & 4 & 1& 5&\ydiagram{1}\\
        (0,2) & 10 & 2& 14&\ydiagram{2}\\
        (0,3) & 18 & 2 & 30&\ydiagram{3}\\
        (2,0) & 6 & 2 & 10&\ydiagram{1,1}\\
        (2,1) & 12 & 3 & 35&\ydiagram{2,1}\\
        (2,2) & 20 & 5 & 81&\ydiagram{3,1}\\
        (4,0) & 16 & 3 & 35&\ydiagram{2,2}\\
        \hline
    \end{tabular}
    \caption{Some $\mathfrak{sp}(4)$ representations with center quantum number $+1$}
    \label{tab:sp4 reps}
\end{table}

\begin{equation}
    Q_{IS}(\lambda_i, I) = \sum_i (\delta\Delta_i-\delta\Delta_i^{CFT})^2,
\end{equation}
where we choose $\delta\Delta_i = [\Delta_{\partial\phi}-\Delta_{\phi}, \Delta_{\partial\partial\phi}-\Delta_{\phi}, \Delta_{\partial\phi^2}-\Delta_{\phi^2}, \Delta_{\partial\partial\phi^2}-\Delta_{\phi^2}]$, and identify the lightest scalar ($C_2=0$) excitation as $\phi$ in the $\mathcal R=-1$ sector, the lightest scalar excitation in the $\mathcal R=+1$ sector as $\phi^2$, the lightest state in $(4,-,+)$ sector to be $\partial_\mu \phi$, etc. $\delta\Delta^{CFT}_i=[1.0, 2.0, 1.0,2.0]$. Minimizing $Q_{IS}$ at $I=2$, we obtain the critical point parameters:
\begin{equation}
    V_{0,1} = 1.600,\qquad V_{1,0} =0.011,\qquad h_x = 5.366,\label{eq:Ising_para_ED}
\end{equation}
with $V_{0,0}=1$ set as the interaction energy scale, and we find that the scalar primaries have scaling dimension:
\begin{equation}
\Delta_\phi=1.38\qquad\Delta_{\phi^2}=2.92\qquad \Delta_{\phi^3} = 4.62,
\end{equation}
which shows a free-scalar spectrum with deviation of order 10\% . We plot the cost function $Q_{IS}$ along the transition in figure \ref{fig:spec_IS}(a).

In the simulation, we may reduce the dimension of the Hilbert space by choosing the total $\mathfrak{so}(5)$ Cartan quantum numbers to be $(0,0)$, the degeneracy can be interpreted as the $\mathfrak{so}(5)$ representation degeneracy at weight $(0,0)$, which is listed in table \ref{tab:sp4 reps}.

For symmetric tensor representations of $\mathfrak{so}(5)$, namely the $(0,q)_5$ representations with $C_2=0,4,10,18,...$, exhibit good conformal level spacing, that we plot in figure \ref{fig:spec_IS}(b). The conformal tower of $\phi,\phi^2,\phi^3$ can be identified, which shows a free-scalar CFT spectrum at the critical point.

We may also try to match the primaries in antisymmetric representations or mixed representations, in the free-scalar CFT, for $(2,0)_5$ there is $\partial_{[\mu}\phi \partial_{\nu]}\phi^2$ with $\Delta=6.5$, for $(2,1)_5$, $\partial_{[\rho}\phi \partial_{\mu]}\partial_\nu\phi$ with scaling dimension $\Delta=6$ which is a descendant of $T_{\mu\nu}$. These states may be identified in figure \ref{fig:spec_IS}(c) and table \ref{tab: full spec IS}. However, there are states that have lower energy level than those in the same sector. Moreover, the first parity odd state should be $\epsilon^{\mu\nu\rho\beta\gamma}\partial_\mu\phi \partial_\nu\partial_\alpha \phi \partial_\rho \phi^2$ with $C_2=12,\Delta=10$, and numerically, parity odd states appear in lower scaling dimensions.

\section{Yang--Lee CFT}
We add a term breaking the $\mathcal R$ inversion symmetry with imaginary coefficients, such that the combination of inversion and time-reversal symmetry is preserved, resulting in discrete symmetries:
\begin{equation}
\begin{split}
   \mathcal R\mathcal T:& \Psi_{\mathbf N}\rightarrow\sigma^x \Psi_{\mathbf N},\qquad i\rightarrow-i\\
   \mathcal P:& \Psi_{\mathbf N}\rightarrow i\sigma^y \Psi_{\mathbf{N}}^\dagger.
\end{split}
\end{equation}
The Hamiltonian reads:
\begin{equation}
\begin{split}
    H &= \sum_{\mathbf{N_1},\mathbf{N_2},\mathbf{N_3},\mathbf{N_4}} V_{\mathbf{N_1},\mathbf{N_2},\mathbf{N_3},\mathbf{N_4}}(::\Psi^\dagger_{\mathbf{N_1}}\Psi_{\mathbf{N_4}}\Psi^\dagger_{\mathbf{N_2}}\Psi_{\mathbf{N_3}}::+::\Psi^\dagger_{\mathbf{N_1}}\sigma^z\Psi_{\mathbf{N_4}}\Psi^\dagger_{\mathbf{N_2}}\sigma^z\Psi_{\mathbf{N_3}}::)\\
    &- h_x \sum_{\mathbf{N}}\Psi^\dagger_{\mathbf N} \sigma^x\Psi_{\mathbf N} + i h_z\sum_{\mathbf{N}}\Psi^\dagger_{\mathbf N} \sigma^z\Psi_{\mathbf N}.\label{eq: YL}
\end{split}
\end{equation}
The notation is the same as the Ising model, i.e. two component fermions in the $(I,0)_5$ representation of the $\mathfrak{sp}(4)$ algebra. As $H^T = H$, the Hermitian condition is replaced by $\mathcal RH^\dagger\mathcal R=H$.

\begin{figure}[t]
    \centering
        \includegraphics[width=1.0\linewidth]{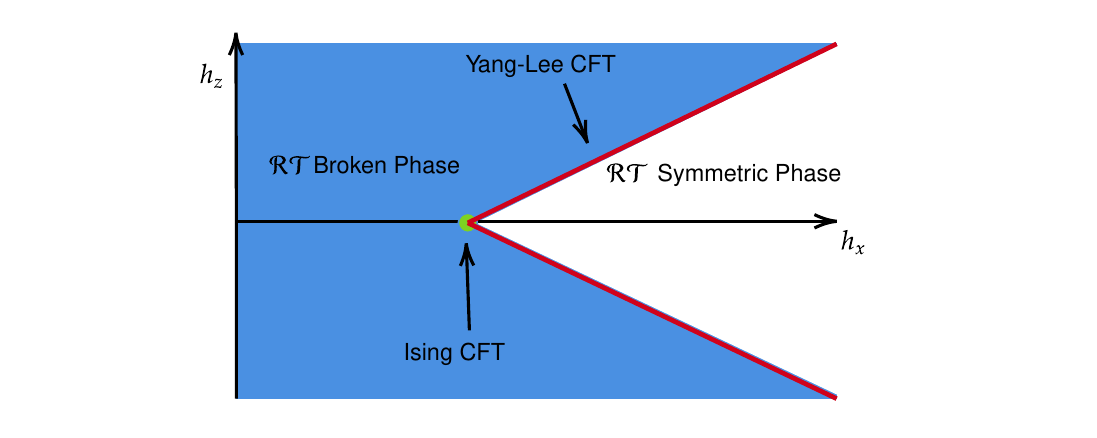}
    \caption{Phase diagram of the Yang--Lee models, adapted from Cruz et al.}
    \label{fig:YL phase}
\end{figure}

Previous works \cite{Cruz_25,Cardy_85,Gehlen_1991,vongehlen_94,Uzelac_81,Castro-Alvaredo_2009,miro_25,fan_25} in $d=1,2,3$ give a phase diagram of Yang--Lee model constructed from spin chains, fuzzy sphere and icositetrachoron, which is illustrated in figure \ref{fig:YL phase}. Starting from the $h_x>h_{x,c}$ paramagnetic phase of the quantum Ising model, and switching on a small $h_z$, the ground state stays $\mathcal R\mathcal T$ symmetric and the energies are real. As increasing $h_z$, the continuation of the Ising ground state and the $\mathcal R=-1$ ground state merge in energy level and goes into complex conjugate pairs, the $\mathcal R\mathcal T$ symmetry operation exchange the two states (in appendix \ref{app: two level model} we give a two-level toy model for the energy level merging). Before the merging point there is a critical value $h_z^*$ that realizes the Yang--Lee CFT.

The Yang--Lee CFT corresponds to the IR fixed point of the Lagrangian:
\begin{equation}
    \mathcal L = \frac{1}{2}(\nabla\phi)^2 +ig\phi^3,
\end{equation}
where the $\mathcal R\mathcal T$ symmetry acts as $\phi\rightarrow-\phi, i\rightarrow-i$. There is a relevant deformation in this IR CFT given by $i\phi$, and the equation of motion $\Box \phi =3ig \phi^2$ removes $\phi^2$ from the primary spectrum. From the $\epsilon$-expansion\cite{PhysRevD.103.116024,Schnetz_25wtu}  $\phi,\phi^3$ and the lowest spin-4 primary $Q_{\mu\nu\rho\sigma}$ have scaling dimensions:
\begin{equation}
    \Delta_{\phi}=1.43, \qquad\Delta_{i\phi^3} =5.70,\qquad \Delta_Q =6.91.
    \label{eq: YL_eps_exp}
\end{equation}

As the conventional positive semi-definite bootstrap method\cite{Simmons-Duffin_2015qma} only applies for unitary theories, one may use the Gliozzi truncated bootstrap method for Yang-Lee CFT\footnote{By inputing the fusion rule and truncate the conformal block expansion to $N$ low lying operators, then the first M Taylor coefficients of the crossing equation forms a $M\times N$ linear system with $M\geq N$, the condition of having a solution translates into all it's $N\times N$ minors are zero.}\cite{Gliozzi_13ysa,Hikami_17}, leading to the scaling dimensions:
\begin{equation}
    \Delta_{\phi}=1.44,\qquad \Delta_Q =6.91.
    \label{eq: YL_CB}
\end{equation}

\begin{figure}[t]
    \centering
        \begin{minipage}[t]{0.45\linewidth}
        \stackinset{l}{2mm}{t}{2mm}{\textbf{a.}}{
        \includegraphics[width=\linewidth]{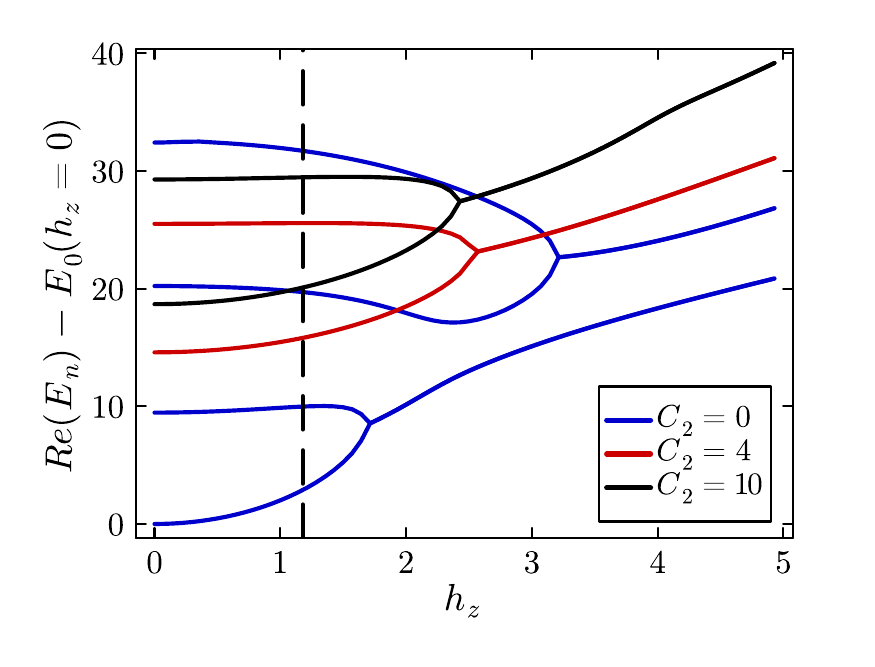}
        }
    \end{minipage}
    \begin{minipage}[t]{0.45\linewidth}
    \stackinset{l}{2mm}{t}{2mm}{\textbf{b.}}{
        \includegraphics[width=\linewidth]{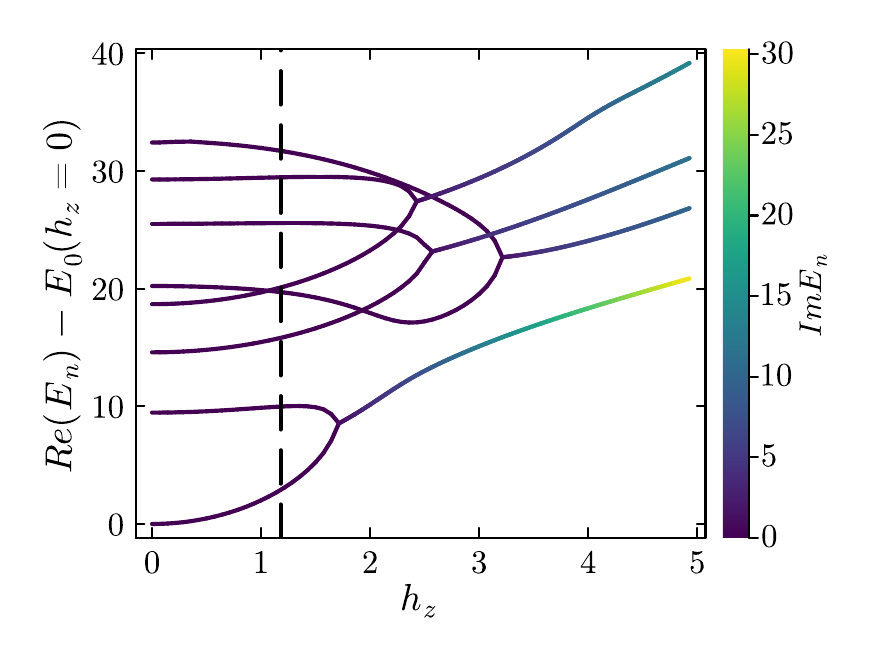}
        }
    \end{minipage} 
    \caption{(a) The energy levels $\text{Re}E_n(h_z)$ of (\ref{eq: YL}). (b) The color denotes the imaginary part of the energies as function of $h_z$, the merging point is $h_z^{(merg.)}\simeq 1.7$, the vertical line denotes the critical point $h_z^*<h_z^{(merg.)}$, which will be determined later. The parameters are $V_{0,1} = 1.862, V_{1,0} =0.061, h_x = 8.913$.}
    \label{fig:Enrg vs hz}
\end{figure}

\begin{figure}[t]
    \centering
        \begin{minipage}[t]{0.45\linewidth}
        \stackinset{l}{2mm}{t}{2mm}{\textbf{a.}}{
        \includegraphics[width=\linewidth]{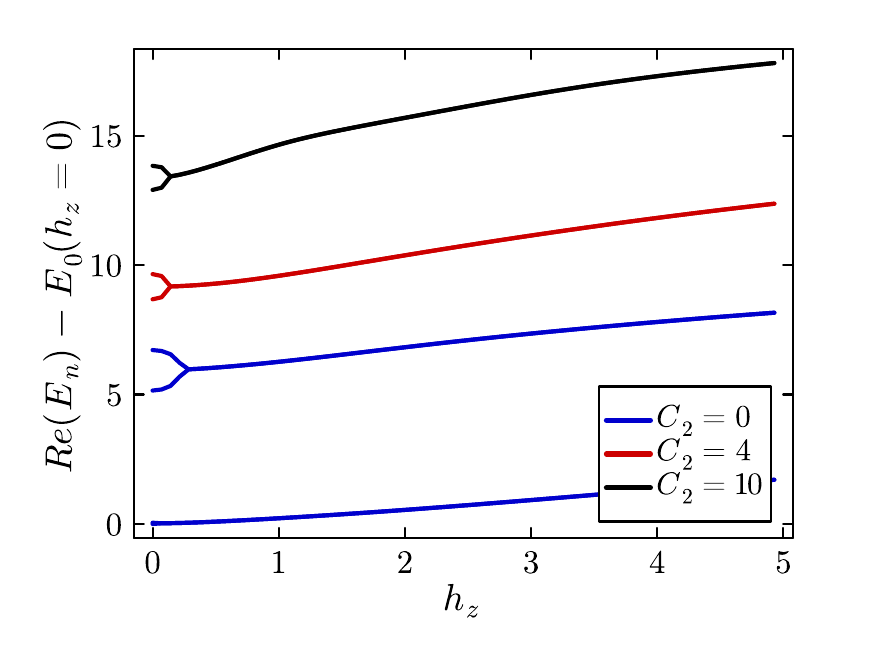}
        }
    \end{minipage}
    \begin{minipage}[t]{0.45\linewidth}
    \stackinset{l}{2mm}{t}{2mm}{\textbf{b.}}{
        \includegraphics[width=\linewidth]{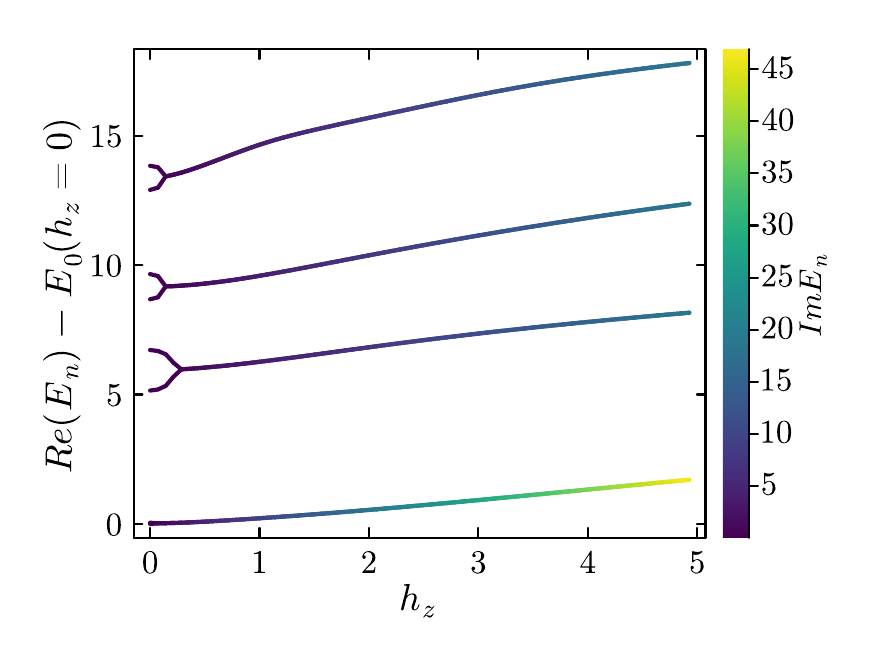}
        }
    \end{minipage} 
    \caption{The same plot as figure \ref{fig:Enrg vs hz}, with different parameters $V_{0,1} = 1.862, V_{1,0} =0.061, h_x = 3.000$, for which $h_z=0$ is the Ising ferromagnetic phase (namely for $h_x$ less than critical value).}
    \label{fig:Enrg vs hz small hx}
\end{figure}

On the fuzzy 4-sphere, we observe behavior similar to $d=1,2,3$. As we start from the Ising paramagnetic phase and turn on $h_z$, at $h_z=h_z^{(merg.)}$, energy levels merge into complex conjugate pairs, see figure \ref{fig:Enrg vs hz}. If we start from the Ising ferromagnetic phase, the ground states in the $\mathcal R=\pm1$ sector merge once we open $h_z$, see figure \ref{fig:Enrg vs hz small hx}.

We search for a critical point by searching for the conformal tower formed by the $\phi$ operator. From the $\epsilon$-expansion result, other than $T_{\mu\nu}$ there are no additional primaries below $\Delta_\phi +4$, so the low energy states are solely the primaries $\phi,T$ and their conformal descendants. We first identify the second state in $(C_2,\mathcal P)=(10,+1)$ sector as $T_{\mu\nu}$ and set $\Delta_T=5$ to fix the $v/R$ normalization, and define a cost function $Q_{YL}$ to quantify the deviation of the spectrum from integer spacing. In this case we add weights to the operators, which make the cost function more sensitive to operators with lower scaling dimensions:
\begin{equation}
\begin{split}
    Q_{YL}(\lambda_i,I) &=\frac{\sum_i W_i(\Delta_i-\Delta_i^{CFT})^2}{\sum_i W_i}, \\
    W_i&= \frac{1}{\Delta_i},
\end{split}
\end{equation}
with
\begin{equation}
\begin{split}
    \Delta_i&=[\Delta_{\partial\phi}, \Delta_{\Box\phi}, \Delta_{\partial\partial\phi}, \Delta_{\partial\Box\phi},  \Delta_{\Box^2\phi}, \Delta_{\partial\partial\Box\phi}],\\
    \Delta_i^{CFT}&=[\Delta_{\phi}+1, \Delta_{\phi}+2, \Delta_{\phi}+2, \Delta_{\phi}+3,  \Delta_{\phi}+4, \Delta_{\phi}+4].
\end{split}
\end{equation}

\begin{figure}
    \centering
        \begin{minipage}[t]{0.45\linewidth}
        \stackinset{l}{2mm}{t}{2mm}{\textbf{a.}}{
        \includegraphics[width=\linewidth]{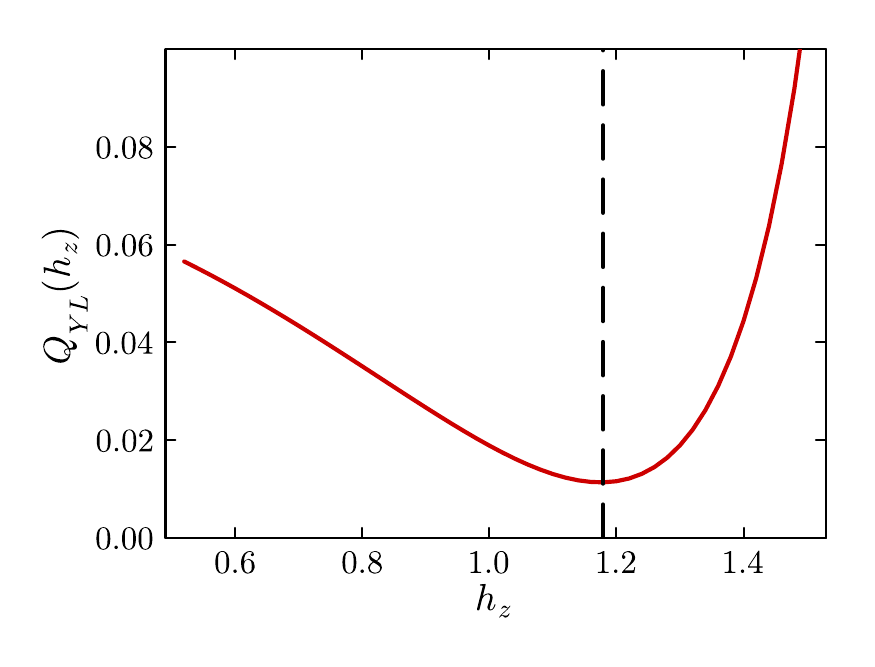}
        }
    \end{minipage}
    \begin{minipage}[t]{0.45\linewidth}
    \stackinset{l}{2mm}{t}{2mm}{\textbf{b.}}{
        \includegraphics[width=\linewidth]{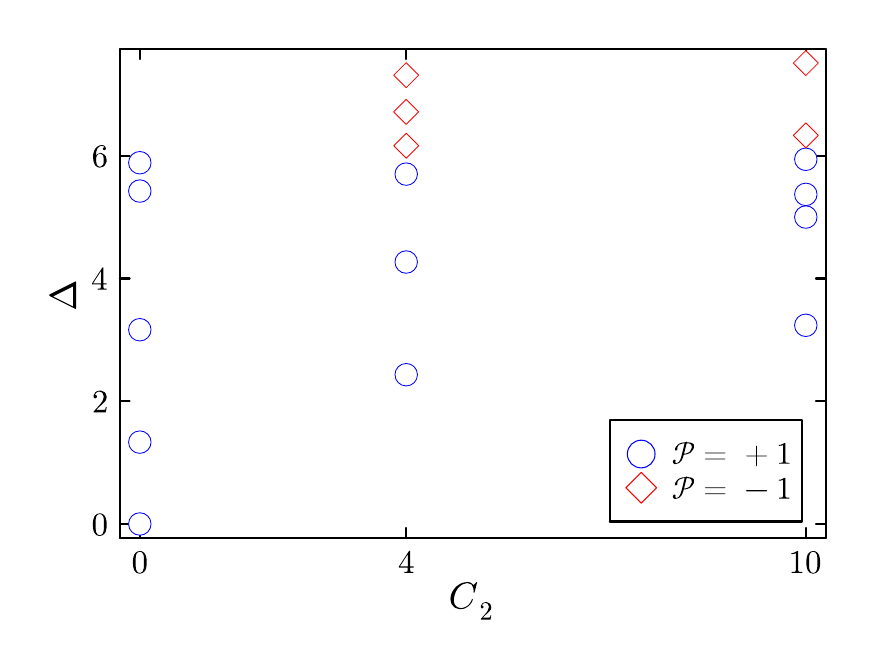}
        }
    \end{minipage} 
    \begin{minipage}[t]{0.7\linewidth}
    \stackinset{l}{2mm}{t}{2mm}{\textbf{c.}}{
        \includegraphics[width=\linewidth]{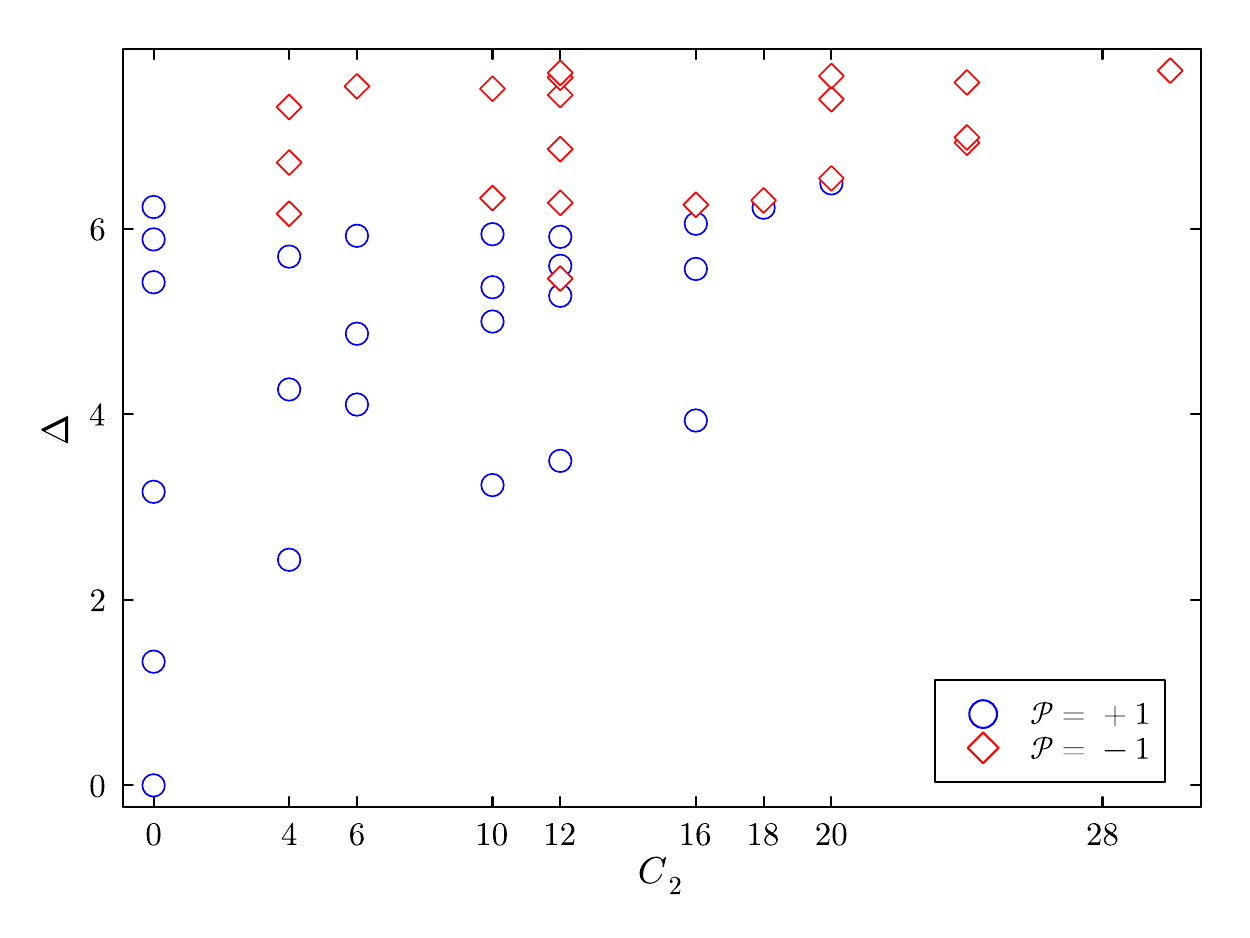}
        }
    \end{minipage}
    \caption{(a) The Yang--Lee cost function as a function of $h_z$, where $V_{0,1}, V_{1,0}, h_x$ are chosen as the optimized value. (b) The critical spectrum, with symmetric tensor representations $(0,q)_5$ at $I=2$. (c) The full spectrum at $I=2$.}
    \label{fig:spec crit YL}
\end{figure}

$Q_{YL}$ is blind to the conformal data of $\Delta_\phi,\Delta_{\phi^3}$\footnote{Other than the assumptions of scaling dimension ordering.}. We found a choice of parameters at $I=2,N_e=10$ that minimize $Q_{YL}$:
\begin{equation}
    V_{0,1} = 1.862,\qquad V_{1,0} =0.061, \qquad h_x = 8.913,\qquad h_z =1.179,\label{eq:YL_para_ED}
\end{equation}
again having set $V_{0,0}=1$. We plot the $Q_{YL}$ along the transition in figure \ref{fig:spec crit YL}(a). From the spectrum at the critical point, see figure \ref{fig:spec crit YL}(b),(c) and table \ref{tab: full spec YL}, we obtain the scaling dimension of $\phi$ operator and of an extra scalar primary, which is identified as $\phi^3$,
\begin{equation}
    \Delta_\phi = 1.33,\qquad \Delta_{\phi^3} = 5.88.
\end{equation}
The results above at $I=2,N_e=10$ are near the $\epsilon$-expansion value (\ref{eq: YL_eps_exp}).

\begin{figure}[t]
    \centering
    \includegraphics[width=0.7\linewidth]{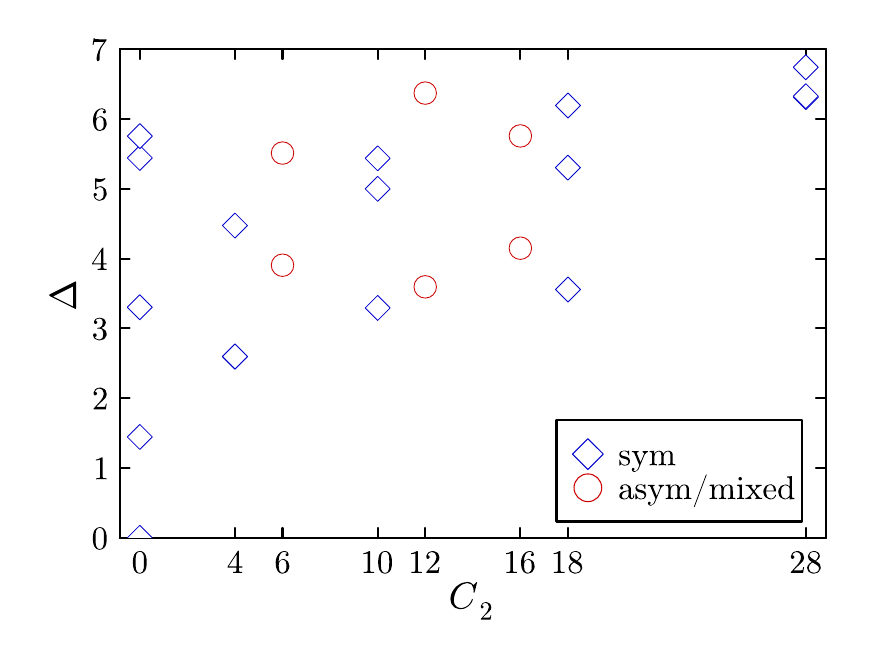}
    \caption{The Yang--Lee spectrum at $I=3,N_e=20$ from DMRG.}
    \label{fig:YL spec DMRG}
\end{figure}

To confirm the above result, we compute the low lying spectrum at $I=3,N_e=20$ using DMRG method. As a non-Hermitian system, the eigenstates are not generically orthogonal, therefore we use an alternative method to define projections in computing the excited states\cite{Carlon:1999eh}, see appendix \ref{app: non_H DMRG}. As the UV critical point may have large finite size drift, we search for the critical values independently in $I=3,N_e=20$ by minimizing the cost function $Q_{YL}$. We perform DMRG using the symmetric projector and $H\rightarrow H + w C_2$ with $w=20$ to raise the energy levels of higher $C_2$ sector. (this helps us to increase the efficiency and accuracy when targeting specific states with $C_2=0,4,10$). We used $2.38\times10^6$ CPU seconds to search for the minimum, and the result is:
\begin{equation}
    V_{0,1} = 2.351,\qquad V_{1,0} =0.060, \qquad h_x = 19.221,\qquad h_z =1.004,\label{eq: YL_para_DMRG}
\end{equation}
where the spectrum cost is $Q_{YL}^*= 0.013$. We further compute the spectrum using DMRG with maximal link dimension $\chi=4000$. The $C_2$ quantum numbers numerically deviate from the quantized values of order $10^{-2}$, and we show the spectrum in figure \ref{fig:YL spec DMRG} and table \ref{tab: full spec YL}. The scaling dimension of $\phi, \phi^3$ and comparison with existing results is shown in table \ref{tab: scaling_dim_compare}. The lightest spin-4 operator $Q_{\mu\nu\rho\sigma}$ may be identified in $C_2=28$ sector. However, the conformal tower structure is not obvious for symmetric tensors with spin larger than 3. There is an analogous situation on fuzzy 2-sphere, when one side of the phase transition is a topological phase, the conformal structure is buried in roton states, which appears below the unitarity bound\cite{Zhou_25}.

\begin{table}[t]
    \centering
    \begin{tabular}{c|c|c|c|c}
    \hline
         &  $\epsilon-$expansion & CB & Fuzzy $\mathbb S^4,N_e=10$& Fuzzy $\mathbb S^4,N_e=20$\\
        \hline
        $\Delta_\phi$ & 1.43 & 1.44 &1.33 & \textcolor{blue}{1.44}$\pm 0.01$\\
         $\Delta_{\phi^3}$ & 5.70 &-- & 5.88 & \textcolor{blue}{5.75}$\pm 0.02$\\
         \hline
    \end{tabular}
    \caption{Comparison of scaling dimensions from various methods, the error bar of $N_e=20$ result comes from comparing the result of $\chi=4000$ and $\chi=6000$.}
    \label{tab: scaling_dim_compare}
\end{table}

\section{Discussions}\label{sec:discussion}

In this work we constructed the critical Ising model and the critical Yang--Lee model on a fuzzy $\mathbb S^4$, and realized a the 5d free-scalar CFT and the 5d Yang--Lee CFT on $\mathbb S^4\times\mathbb R$. To be specific, we show continuous phase transitions and conformal spectrum in the symmetric representation sectors with small spin. We independently determine the Yang--Lee scaling dimensions of the primary scalars $\phi,\phi^3$.

It is surprising that at $I=2$ $N_e=10$, the model can give a conformal spectrum with fine tuning, however extending the model to a larger size is technically hard. The ground state degeneracy scales with $I^3$, which make the study of finite size scaling limited. Also, in four spatial dimensions, the entanglement entropy of low lying states is expected to grow as $S\sim R^3$, which results in a fast growth of the link dimension needed by MPS approximation in DMRG method.

Despite these positive results, for anti-symmetric $\mathfrak{so}(5)$ and mixed representation sectors, there are extra low energy states for both free-scalar model and Yang--Lee model. From the comparison of $I=2,3$ results, some of the low-lying states are persistent as increasing $I$, which may be interpreted as a sector decoupled with the CFT we identified.

One could compute the OPE coefficients for 5d Yang-Lee model, for example from $n^z \sim c_\phi \phi +c_\phi^{(1)} \partial_\mu\phi+...$, using $\frac{\langle \phi |n^z(\Omega)|\phi \rangle}{\langle \phi |n^z(\Omega)|I \rangle}=C_{\phi\phi\phi}+O(R^{-1})$. To subtract the $O(R^{-1})$ contribution, one may first find method to determine $R(I)$.

\acknowledgments{Jiangyuan Qian would like to thank Wei Zhu for useful discussions, and Ofer Aharony for useful discussions and helps on organizing the paper. The numerical computations have used a public available Julia package FuzzifiED\cite{Zhou_25pkg}. This work was supported in part by ISF grant no.2159/22, by Simons Foundation grant 994296 (Simons Collaboration on Confinement and QCD Strings), by the Minerva foundation with funding from the Federal German Ministry for Education and Research, and by the German Research Foundation through a German-Israeli Project Cooperation (DIP) grant “Holography and the Swampland”. The numerical computations were performed using the WEXAC high-performance computing cluster at Weizmann Institute of Science.
}

\appendix

\section{\texorpdfstring{$\mathfrak{sp}(4)\simeq \mathfrak{so}(5)$ representations}{sp4 rep.}}\label{app: sp4_rep}
Here we take the Cartan--Weyl normalization, with:
\begin{equation}
\begin{split}
    \text{Tr}(E_{\alpha}E_{-\alpha})&= \frac{2}{|\alpha|^2}\\
    \text{Tr}(H^iH^j) &= \delta^{ij}.
\end{split}
\end{equation}
Every root with it's adjoint and a Cartan element form a $\mathfrak{su}(2)$ sub-algebra:
$$
h^i = \frac{2\alpha\cdot H}{\alpha^2}, e^i = E_{\alpha^i}, f^i = E_{-\alpha^i}.
$$
In the following text, we will use the $\mathfrak{sp}(4)$ notation of $\mathfrak{so}(5)$ algebra, it has two simple roots and, the Cartan matrix is given by:
\begin{equation}
    A=\begin{pmatrix}
        2& -1 \\-2 & 2
    \end{pmatrix},
\end{equation}
where each row denotes the $\mathfrak{su}(2)$ quantum numbers of a simple root. We use these $\mathfrak{su}(2)$ quantum numbers of the highest weight state to label the state, for instance $(1,0)_5$ labels the $\mathbf 4$ spinor representation of $\mathfrak{so}(5)$, $(0,1)_5$ labels the $\mathbf 5$ vector representation. On the fuzzy 4-sphere, the LLL projection restricts the fermionic orbital to the $(I,0)_5$ representation, which can be constructed by a symmetric product of $I$ copies of spinor representation:
\begin{equation}
    \Psi_{i_1,...i_I} = S_{i_1,...,i_I} \psi^{i_1}...\psi^{i_I},
\end{equation}
As a bosonic state is totally symmetric, by introducing bosonic operators $a^\dagger_i$ ($i=1,2,3,4$) that transform in the $\mathbf 4$ representation and imposing $N=\sum_{i=1}^4a^\dagger_i a_i=I$, one can obtain the $(I,0)_5$ representation. This Schwinger boson trick is not only useful in providing a basis and allowing computation of representation matrices, it also allows to compute a special class of CG coefficients. We first show the representation matrix:
\begin{equation}
    D(A) = \sum_{i,j}a^\dagger _i A_{ij}a_j,
\end{equation}
where $A_{ij}$ is the matrix element in the fundamental representation $\textbf{4}$. One can check that $[D(A),D(B)]=D([A,B])$, and $[D(A),N]=0$. The generators are:
\begin{equation}
    \begin{split}
        E_{\alpha_1} &= a^\dagger_1 a_2 - a^\dagger_4 a_3,\\
        E_{\alpha_2} &= a^\dagger_2 a_4,\\
        E_{\alpha_1+ \alpha_2} &= a^\dagger_1 a_4+ a^\dagger_2 a_3,\\
        E_{2\alpha_1+ \alpha_2} &= a^\dagger_1 a_3,
    \end{split}
\end{equation}
and the two Cartan elements:
\begin{equation}
\begin{split}
     h^1 &= a_1^\dagger a_1 -a_2^\dagger a_2-a_3^\dagger a_3 + a_4^\dagger a_4,\\
    h^2 &= a_2^\dagger a_2 -a_4^\dagger a_4.
\end{split}
\end{equation}
After a little algebra one can confirm that:
\begin{equation}
    \begin{split}
       [ h^i , e^j ] &= A_{ji} e^j\qquad \text{(no sum)}\\
       [e^j, f^j] & = h^j.
    \end{split}
\end{equation}
As the number operator $N$ becomes a Casimir operator, the $(I,0)_5$ representation lies in the $N=I$ sector of the Hilbert space. The dimension of such a space is equal to the ways of partitioning $I$ to four non-negative integers with $d(I,0) ={I+3\choose 3}$. It is also useful to introduce a bipartite system, the boson creation operators are denoted as $a_i^\dagger, b_i^\dagger$ for two subsystems, and one can construct a generically reducible representation in this Hilbert space:
\begin{equation}
    \begin{split}
        D(E_\alpha) = D_A(E_\alpha) \otimes D_B(I)+ D_A(I) \otimes D_B(E_\alpha).
    \end{split}
\end{equation}
One can enumerate the irreducible representations by constructing highest weight states, which are states annihilated by:
\begin{equation}
    \begin{split}
        E_{\alpha_1} &= a^\dagger_1 a_2 - a^\dagger_4 a_3+b^\dagger_1 b_2 - b^\dagger_4 b_3,\\
        E_{\alpha_2} &= a^\dagger_2 a_4+ b^\dagger_2 b_4.\\
    \end{split}
\end{equation}
The simplest classes of states are:
\begin{equation}
    (a_1^\dagger)^{I_a} (b_1^\dagger)^{I_b}|0\rangle.
\end{equation}
Denote the operator $D = a_1^\dagger b_3^\dagger+ a_2^\dagger b_4^\dagger-a_3^\dagger b_1^\dagger-a_4^\dagger b_2^\dagger$ as the symplectic form which commutes with all the generators, another class of highest weight states are given by:
\begin{equation}
    D^r (a_1^\dagger)^{I_a-r} (b_1^\dagger)^{I_b-r}|0\rangle,
\end{equation}
and we can also notice an anti-symmetric operator:
\begin{equation}
    \Lambda = a^\dagger_2 b^\dagger_1 - a^\dagger_1b^\dagger_2.
\end{equation}
Intuitively applying $\Lambda$ adds a column with two boxes to the Young tableaux. It commutes with $E_{\alpha_1}, E_{\alpha_2}$, and the (unnormalized) highest weight state has the form:
\begin{equation}
  D^r\Lambda^k (a_1^\dagger)^{I_a-r-k} (b_1^\dagger)^{I_b-r-k}|0\rangle.
\end{equation}
Practically, we construct a representation in the following way:
\begin{itemize}
    \item write the (normalized) highest weight state on the bipartite system,
    \item apply $D(E_{-\alpha_1})$ and $D(E_{-\alpha_2})$ to get the next level,
    \item orthogonalize, and normalize the new vectors,
    \item remove zero vectors from the new set,
    \item if the number of new basis states is nonzero, go back to bullet 2.
\end{itemize}
In this way we get the representations $(I_a+I_b-2k-2r,k)_5$ for $I_a, I_b,k,r\geq0$, and importantly the CG coefficient of
\begin{equation}
    (I_a,0)_5\otimes (I_b,0)_5\rightarrow(I_a+I_b-2k-2r,k)_5
\end{equation}
can be read-off by projecting the bipartite state vectors to the direct product state vector:
\begin{equation}
   C_{\mathbf N \mathbf {\tilde{N}}}^{(p,q,i)} = \langle \mathbf N ,\mathbf {\tilde{N}}|(p,q)_5,i\rangle,
\end{equation}
where $\mathbf N ,\mathbf {\tilde{N}}$ denote the occupation number vector in the two "subsystems", and $i=1,...,d(p,q)$ labels states in the $(p,q)_5$ representation.

We also construct the second Casimir operator. In a general basis, it has the form:
\begin{equation}
    C_2 = K_{ab}J^a J^b,
\end{equation}
where $K_{ab}$ is the positive definite Killing form. In the Cartan--Weyl basis, it takes the form:
\begin{equation}
\begin{split}
      C_2 &= \sum_{i=1}^r H^i H^i + \sum_{\alpha^i\in\Delta^+} \frac{|\alpha^i|^2}{2}(E_{\alpha^i} E_{-\alpha^i}+E_{-\alpha^i} E_{\alpha^i}),\\
    C_2 &= \sum_{i=1}^r H^i H^i + \sum_{\alpha^i\in\Delta^+} \left(\alpha^i\cdot H + |\alpha^i|^2 E_{-\alpha_i}E_{\alpha_i}\right),
\end{split}
\end{equation}
where we did not distinguish the Lie algebra and it's representation, and $\Delta^+$ denotes the set of positive roots. 

For the representation $(p,q)_5$ the highest weight can be written as $\lambda=p \omega_1+q\omega_2$, $E_{\alpha_i}|\lambda\rangle=0$, Defining the inner product of roots and weights: $\sum_{i=1}^2\alpha^i\beta^j=(\alpha,\beta)$ (where $\alpha^i,\beta^i$ are in the symmetric basis) we are left with:
\begin{equation}
\begin{split}
     C_2 (p,q) = (\lambda,\lambda)+\sum_{\alpha^i\in\Delta^+} (\alpha^i,\lambda),\\
\end{split}
\end{equation}
where $G_{ij}$ is the metric on the basis of fundamental weights
\begin{equation}
   G_{ij}= (\omega_i, \omega_j)=\frac{1}{2}\begin{pmatrix}
        1 & 1\\
        1 & 2
    \end{pmatrix}
\end{equation}
As the positive roots are written as:
\begin{equation}
    \begin{split}
        \alpha_1 &= 2\omega_1 - \omega_2\\
        \alpha_2 &= -2\omega_1 + 2\omega_2\\
        \alpha_1+\alpha_2 &= \omega_2\\
        2\alpha_1+\alpha_2 &= 2\omega_1 ,\\
    \end{split}
\end{equation}
we find:
\begin{equation}
\begin{split}
     C_2 (p,q) = \frac{1}{2}\begin{pmatrix}
         p&q
     \end{pmatrix} \begin{pmatrix}
         1&1\\
         1&2
     \end{pmatrix}
     \begin{pmatrix}
         p\\
         q
     \end{pmatrix}+ \frac{1}{2}\begin{pmatrix}
         2&2
     \end{pmatrix}\begin{pmatrix}
         1&1\\
         1&2
     \end{pmatrix} \begin{pmatrix}
         p\\
         q
     \end{pmatrix},
\end{split}
\end{equation}
\begin{equation}
\begin{split}
     C_2 (p,q) = \frac{1}{2}\left(p^2 + 2pq + 2q^2 +4p +6q\right).
\end{split}
\end{equation}
 The $\mathbb Z_{2}$ central elements $-I_{\mathbf 4\times \mathbf 4}$ act on a representation $(p,q)_5$ as the center quantum number $(-1)^{p+2q}=(-1)^{p}$, since it can be obtained from the $I_a+I_b=p+2q$ tensor product representation which has center representation $(-1)^{p}$. 

For a many-body fermionic system, the $\mathrm{SO}(5)$ symmetry operator can be written as:
\begin{equation}
    \hat{\mathbf{G}} =\sum_{i,j=1}^{d(I,0)}\sum_f c^\dagger_{if} D(G)_{ij}^{(I,0)}c_{jf},
\end{equation}
where $D(G)_{ij}^{(I,0)}$ is the $(I,0)_5$ representation matrix. One can easily check it forms a closed algebra. Plugging $\hat{\mathbf{G}}$ into the form of the second Casimir, one can construct the many-body $\hat C_2$ operator, as we computed in the main text.
\section{Spectrum}\label{app:spectrum}
We listed the Ising model spectrum for $I=2$ with parameters (\ref{eq:Ising_para_ED}) in table \ref{tab: full spec IS}, and the Yang-Lee spectrum for $I=2,3$ with parameters (\ref{eq:YL_para_ED}), (\ref{eq: YL_para_DMRG}) respectively, in table \ref{tab: full spec YL}.

\begin{table*}[t]
   \setlength{\tabcolsep}{12pt}
   \renewcommand{\arraystretch}{1.5}
    \centering
    \caption{The Ising spectrum at $I=2$, the low-lying symmetric representations can be identified}
    \label{tab: full spec IS}
    \begin{tabular}[t]{ccccc}
    \hline
       $C_2$  & $\Delta$ & $\mathcal R$ & $\mathcal P$ & $\text{op.}$\\
    \hline
  0 & 0.0 & 1 & 1 & $\mathbb I $ \\
  0 & 1.38237 & -1 & 1 & $\phi$ \\
  4 & 2.42872 & -1 & 1 & $\partial_\mu \phi$ \\
  0 & 2.92365 & 1 & 1 & $\phi^2$ \\
  10 & 3.53339 & -1 & 1 & $\partial_\mu\partial_\nu \phi$ \\
  12 & 3.91583 & -1 & 1 &  \\
  4 & 4.00122 & 1 & 1 & $\partial_\mu\phi^2$ \\
  16 & 4.41252 & -1 & 1 &  \\
  6 & 4.45266 & -1 & 1 &  \\
  0 & 4.61928 & -1 & 1 & $\phi^3$ \\
  6 & 4.75118 & 1 & 1 &  \\
  10 & 5.0 & 1 & 1 & $T_{\mu\nu}$ \\
  0 & 5.15042 & -1 & 1 &  \\
  10 & 5.28802 & 1 & 1 & $\partial_\nu\partial_\mu\phi^2$ \\
  12 & 5.41977 & 1 & 1 &  \\
  12 & 5.44212 & 1 & 1 &  \\
  4 & 5.71641 & 1 & 1 &  \\
  0 & 5.77053 & 1 & 1 &  \\
  16 & 5.8828 & 1 & 1 &  \\
  4 & 5.9578 & -1 & 1 & $\partial_\mu \phi^3$ \\
    \hline
    \end{tabular}
\!\!\!
    \begin{tabular}[t]{ccccc}
    \hline
       $C_2$  & $\Delta$ & $\mathcal R$ & $\mathcal P$ & $\text{op.}$\\
    \hline
  12 & 5.6033 & 1 & -1 &  \\
  4 & 5.74748 & -1 & -1 &  \\
  4 & 6.13324 & 1 & -1 &  \\
  16 & 6.48896 & 1 & -1 &  \\
  12 & 6.50194 & 1 & -1 &  \\
  10 & 6.55054 & 1 & -1 &  \\
  18 & 6.62846 & 1 & -1 &  \\
  4 & 6.8461 & -1 & -1 &  \\
  20 & 6.93953 & 1 & -1 &  \\
  10 & 7.03412 & -1 & -1 &  \\
  12 & 7.1013 & -1 & -1 &  \\
  12 & 7.18269 & -1 & -1 &  \\
  & & & & \\
  & & & & \\
  & & & & \\
  & & & & \\
  & & & & \\
  & & & & \\
  & & & & \\
  & & & & \\
    \hline
    \end{tabular}
\end{table*}

\begin{table*}[t]
   \setlength{\tabcolsep}{17pt}
   \renewcommand{\arraystretch}{1.5}
    \centering
    \caption{left: the Yang--Lee spectrum at $I=2$ using ED, right: the Yang--Lee spectrum at $I=3$ using DMRG.}
    \label{tab: full spec YL}
    \begin{tabular}[t]{cccc}
    \hline
       $C_2$  & $\Delta$ & $\mathcal P$ & $\text{op.}$\\
    \hline
  0 & 0.0 & 1 & $\mathbb I$ \\
  0 & 1.33359 & 1 & $\phi$ \\
  4 & 2.43188 & 1 & $\partial_\mu \phi$ \\
  0 & 3.16456 & 1 & $\Box \phi$ \\
  10 & 3.23705 & 1 & $\partial_\mu \partial_\nu \phi$ \\
  12 & 3.49828 & 1 &  \\
  16 & 3.93324 & 1 &  \\
  6 & 4.10512 & 1 &  \\
  4 & 4.26763 & 1 & $\partial_\mu \Box \phi$  \\
  6 & 4.8687 & 1 &  \\
  10 & 5.0 & 1 &  $T_{\mu\nu}$ \\
  12 & 5.27738 & -1 &  \\
  10 & 5.36974 & 1 &  $\partial_\mu \partial_\nu\Box \phi$\\
  0 & 5.42345 & 1 & $\Box^2 \phi$ \\
  12 & 5.46286 & -1 &  \\
  16 & 5.56652 & 1 &  \\
  12 & 5.60029 & 1 &  \\
  4 & 5.6999 & 1 &  \\
  0 & 5.88527 & 1 & $\phi^3$ \\
   & & & \\
     & & & \\
       & & & \\
    \hline
    \end{tabular}
\!\!\!
    \centering
    \begin{tabular}[t]{cccc}
      \hline
       $C_2$  & $\Delta$ &  $\text{op.}$\\
    \hline
    0 & 0.0 & $\mathbb{I}$ \\
    0 & 1.44527 & $\phi$ \\
    4 & 2.59587 & $\partial_\mu\phi$ \\
    10 & 3.2928 & $\partial_\mu \partial_\nu\phi$ \\
    0 & 3.30289 & $\Box \phi$ \\
    18 & 3.55802 &  \\
    12 & 3.59578 &  \\
    6 & 3.90538 &  \\
    16 & 4.14925 &  \\
    4 & 4.47318 & $\partial_\mu \Box \phi$ \\
    10 & 5.0 & $T_{\mu\nu}$ \\
    18 & 5.30346 &  \\
    10 & 5.43692 & $\partial_\mu \partial_\nu\Box \phi$ \\
    0 & 5.44217 & $\Box ^2 \phi$ \\
    6 & 5.51208 &  \\
    0 & 5.75445 & $\phi^3$ \\
    16 & 5.75752 &  \\
    18 & 6.193 &  \\
    28 & 6.31319 &  \\
    28 & 6.32679 &  \\
    12 & 6.37156 &  \\
    28 & 6.74127 & \\
    \hline
    \end{tabular}
\end{table*}

\section{\texorpdfstring{A two level model exhibiting $\mathcal R \mathcal T$ symmetry breaking}{a two level model exhibit RT sym}}\label{app: two level model}
In this section we give a two level QM model for $\mathcal R \mathcal T$ symmetry breaking. We consider the simplest case: a $\mathbb Z_2$ odd and a $\mathbb Z_2$ even energy level merging. The Hamiltonian and $\mathcal R$ matrix are:
\begin{equation}
    H =\begin{pmatrix}
        H_{11} && H_{12}\\
        H_{12} && H_{22}
    \end{pmatrix}\qquad \mathcal R =\begin{pmatrix}
        +1 && 0\\
       0 && -1
    \end{pmatrix},
\end{equation}
where the $\mathcal R \mathcal T$ symmetry impose $H_{11},H_{22}$ to be real, and $H_{12}$ to be purely imaginary,
\begin{equation}
    H=E_0I +\begin{pmatrix}
        h && i\epsilon\\
        i\epsilon && -h
    \end{pmatrix}.
\end{equation}
We identify $h_z=0$ line where $\mathcal R $ is preserved with $\epsilon=0$. Increasing $\epsilon$, the eigenvalues exhibit:
\begin{itemize}
    \item $\epsilon<h$  $E=E_0\pm \sqrt{h^2 -\epsilon^2}$, a pair of $\mathcal R \mathcal T$ invariant states with real energy,
    \item $\epsilon=h$  $E=E_0$ the pair of states become degenerate,
     \item $\epsilon>h$  $E=E_0\pm i\sqrt{\epsilon^2 -h^2}$ the energies come in complex conjugate pairs and $\mathcal R \mathcal T $ symmetry swaps the energy eigenstate.
\end{itemize}
The eigenstates are given by:
\begin{equation}
    \psi^{\pm} = \mathcal N \begin{pmatrix}
        -i\frac{h \pm\sqrt{h^2-\epsilon^2}}{\epsilon}\\
        1
    \end{pmatrix}\qquad \epsilon<h,
\end{equation}
and one can show that the $\mathcal R \mathcal T$ symmetry action only gives an overall minus sign.

In the $\mathcal R \mathcal T$ breaking regime:
\begin{equation}
    \psi^{\pm} = \mathcal N \begin{pmatrix}
        -i\frac{h \pm i\sqrt{\epsilon^2-h^2}}{\epsilon}\\
        1
    \end{pmatrix}\qquad \epsilon> h,
\end{equation}
and the $\mathcal R \mathcal T$ transformation swaps $\psi^+$ and $\psi^-$. Continuous quantum mechanics models with $\mathcal R \mathcal T$ symmetry are studied in\cite{Bender_98} which has similar behavior.

\section{DMRG algorithm for non-Hermitian symmetric Hamiltonian}\label{app: non_H DMRG}
This section describes the algorithm used for a non-Hermitian symmetric Hamiltonian. We assume that the system is before the merging point and all energies are real and bounded from below (at least for the low-lying states that describe the CFT). In DMRG algorithm for excited states, we need to project out the computed states $|\Psi_i\rangle$, $i=1,...,n$ by lifting their energy, for a Hermitian Hamiltonian this can be done by adding a projection operator to the Hamiltonian:
\begin{equation}
    H\rightarrow H + w \sum_{i=1}^n |\Psi_i\rangle\langle\Psi_i |,
\end{equation}
which lifts the first n energy levels by $w$. In the non-Hermitian case this is not valid as the eigenstates are generically not orthogonal. If the Hamiltonian is symmetric, one can use the following projection:
\begin{equation}
\begin{split}
    H &\rightarrow H + w \Psi_i (S^{-1})_{ij} \Psi _j^T,\\
    S_{ij} &=\Psi _i ^T \Psi _j.
\end{split}
\end{equation}
Notice that for a symmetric Hamiltonian, $\Psi_i^T\Psi_j=0$ if $E_i \neq E_j$, thus the new Hamiltonian lifts the first $1,...,n$ energy levels by $w$ while keeping the rest unchanged.

This allows us to compute the spectrum of a symmetric Hamiltonian even if it is not Hermitian, by using the standard techniques of computing the ground state of the shifted Hamiltonian. In practice, one choose $w$ high enough compare to the energy spacing of the Hamiltonian.

\newpage

\bibliographystyle{JHEP}
\bibliography{library}

\end{document}